\documentclass[aps,prd,a4paper,twocolumn,showpacs,showkeys,preprintnumbers,amsmath,amssymb]{revtex4}
\usepackage{graphicx}
\usepackage{bm}
\usepackage{dcolumn}
\usepackage{color}
\usepackage{pst-plot}
\usepackage{epstopdf}
\usepackage{rotate}
\renewcommand{\texttt}{{}}
\newcommand{\be}{\begin{eqnarray}}
\newcommand{\ee}{\end{eqnarray}}
\usepackage{graphicx}
\begin{document}
\title{Cosmological production of noncommutative black holes} 
\author{Robert B. Mann}
\thanks{Electronic address: rbmann@sciborg.uwaterloo.ca}
\affiliation{Perimeter Institute for Theoretical Physics,
31 Caroline Street North, Waterloo, ON N2L 2Y5, Canada\\
Department of Physics and Astronomy,
University of Waterloo, Waterloo, ON N2L 3G1 Canada}

\author{Piero Nicolini}
\thanks{Electronic address: nicolini@th.physik.uni-frankfurt.de}
\affiliation{Frankfurt Institute for Advanced Studies (FIAS), 
 D-60438  Frankfurt am Main, Germany\\ 
 Institut f\"ur Theoretische Physik, 
J. W.  Goethe Universit\"at, 
 D-60438 Frankfurt am Main, Germany}

\date{\small\today}




\date{\today}





\begin{abstract}
We investigate the pair creation of noncommutative black holes in a background with positive cosmological constant. As a first step we derive the noncommutative geometry inspired Schwarzschild deSitter solution. By varying the mass and the cosmological constant parameters, we find several spacetimes compatible with the new solution: positive mass spacetimes admit one cosmological horizon and two, one or no black hole horizons, while negative mass spacetimes have just a cosmological horizon.  
These new black holes share the properties of the corresponding asymptotically flat solutions, including the non-singular core and thermodynamic stability in the final phase of the evaporation. 
As a second step we determine the action which generates the matter sector of gravitational field equations and we construct instantons describing the pair production of black holes and the other admissible topologies. As a result we find that for current values of the cosmological constant the deSitter background is quantum mechanically stable according to experience.   However positive mass noncommutative black holes and solitons would have plentifully been produced during inflationary times for Planckian values of the cosmological constant. As a special result we find that, in these early epochs of the universe, Planck size black holes production would have been largely disfavoured.  We also find a potential instability for production of negative-mass solitons.
\end{abstract}


\pacs{Valid PACS appear here}
\maketitle

\tableofcontents

\section{Introduction}
Noncommutative geometry inspired black holes (for brevity noncommutative black holes (NCBHs)) are a family of black hole solutions of Einstein equations which incorporate effects of quantum gravity in the short distance/extreme energy regime of the gravitational field of a black hole \cite{review}. The derivation of line elements for NCBHs is based on the possibility of implementing an effective minimal length in general relativity. Instead of embarking in the formulation of the full theory of quantum gravity, it has been shown that the primary effects of manifold quantum fluctuations can be modelled by a non standard form of the energy momentum tensor, while keeping formally unchanged the Einstein tensor in gravity field equations \cite{weak}. Noncommutative geometry is the underlying structure employed to implement, in agreement with the tenets of quantum gravity, a minimal length responsible for delocalization of any point like object and the new form of the energy momentum tensor \cite{NCQFT}.   NCBHs match known black hole solutions at large distances where the presence of the minimal length is negligible, while affording new physics that emerges at small scales. A major result is that for all NCBHs the curvature singularity at the origin is smeared out, being replaced by a regular deSitter core. The presence of a deSitter core discloses further insights about the nature of NCBHs: the centre of NCBHs is a complex turbulent storm-tossed sea which accounts for the seething fabric of spacetime and sustains a Gaussian shaped mass density profile preventing its collapse into a singular distribution. In other words the actual mean effect of manifold fluctuations at short scales is a region characterized by a net repulsive gravitational field.  Energy condition violations at the origin certify that deSitter cores can only be attained in a nonclassical gravity framework.

Another important feature concerns an improved thermodynamics. Indeed, even for neutral NCBHs, the Hawking temperature  reaches a maximum before running a positive heat capacity, cooling down phase towards a zero temperature remnant configuration \cite{ncschw,ncthermo}. As a consequence, according to this scenario quantum back reaction is strongly suppressed in contrast to conventional limits of validity of the semiclassical approximation in the terminal phase of the evaporation.

Both the regularity of the manifold and the improved thermodynamics seem to be model independent results as far as quantum gravity effects are taken into account. Indeed these features of NCBHs are corroborated by analogous results coming from other approaches to quantum gravity \cite{LQBHs,ASGBHs}.  However NCBHs are currently the richest class of quantum gravity black holes since they can be dirty \cite{dirty}, charged \cite{charged}, spinning \cite{spinning} and admit a variety of companion geometries \cite{otherBHs} like traversable wormholes \cite{worm}.  In addition, it has been shown that NCBHs  are connected with a recently proposed ultraviolet complete quantum gravity \cite{Moffat:2010bh}:  modelling manifold fluctuations by means of a noncommutative diffused energy momentum tensor in Einstein gravity turns out to be equivalent to a nonlocal ultraviolet complete quantum gravity produced by a ordinary matter energy momentum tensor \cite{Modesto:2010uh}. As a result for the dual link between the two formulations, NCBHs are solutions of gravity field equations of the ultraviolet complete quantum gravity too.   While the stability of their
interiors remains a subject of ongoing investigation  \cite{stability},
higher dimensional NCBHs \cite{higher,chargedhigher}, due to their attractive properties have been recently taken into account in Monte Carlo simulations as reliable candidate models to describe the conjectured production of microscopic black holes in particle accelerators \cite{Gingrich:2010ed,NCBHpheno}. 

In this paper we want to broaden our perspective. In inflationary epochs, the Universe is well described by the deSitter spacetime, which is known to be stable under classical perturbations \cite{dsstab}. However at quantum mechanical level, the deSitter spacetime exhibits instability due to the spontaneous nucleation of black holes.
The standard semiclassical formalism for pair creation is based on the study of \textit{instantons}. In Euclidean quantum gravity one calculates amplitudes by means of path integrals over various classes of positive defined metrics $g_{ab}$ \cite{Bousso:1996au}
\begin{equation}
\Psi=\int D\left[g_{ab}\right]\ e^{-I_E[g]}.
\label{piaction}
\end{equation}
Usually the evaluation of the path integral is only viable by considering dominant contributions coming from saddle points of $I_E$. Then the path integral is given by a sum over instantons or extrema of the action. In the case of pair production one finds two instantons, one for the background $I_{\mathrm{bg}}$ and one for the object nucleated on the background, $I_{\mathrm{obj}}$. Squaring $\Psi$ one obtains two probability densities whose ratio is the rate of black hole production on inflationary background \cite{Bousso:1996au}, i.e.,
\begin{equation}
\Gamma\sim\frac{\exp\left(-2I_{\mathrm{obj}}\right)}{\exp\left(-2I_{\mathrm{bg}}\right)}.
\label{bhbgrate}
\end{equation}
The background contribution is crucial since it is the positive cosmological constant that supplies the necessary negative potential energy to produce black holes.
 A lot work has been done to calculate the rate of black hole production per unit of volume \cite{Bousso:1996au,dowker,Mann:1995vb,Copsey:2010vb}. However in Einstein gravity the black hole production rate is $\propto e^{-1/\Lambda G}$, an extremely low value  unless the cosmological constant $\Lambda$ approaches the Planck scale.  Thus the
only time when black hole pair creation was possible in our universe was during the inflationary era, when $\Lambda$ was large. In addition the pair produced black holes have Planckian masses and therefore quickly evaporate off.
 In this context we want to investigate the nucleation rate of NCBHs and understand how they affect the quantum (in)stability of the deSitter spacetime. Furthermore NCBHs are longer-lived with respect to conventional black holes, since they do not completely evaporate.  Thus a relevant NCBH production could have potential repercussions for Planck-scale inflation as well as for the production of primordial black holes.
 We remark that NCBHs also efficiently provide reliable scenarios in the semi-classical regime in which the instanton formalism works. This is due to the fact  that noncommutative effects emerge at a length scale $\ell$, which need  not   be the Planck length, but can be treated as a parameter adjustable to the relevant scale at which non-commutative effects set in.

Our paper is organized as follows. In section \ref{line} we derive the noncommutative geometry inspired Schwarzschild-deSitter (NCSchwdS) spacetime for both positive and negative mass parameters, in section \ref{faction} we derive the functional action which leads to the non-standard energy momentum tensor for the solutions  in the previous section; in section \ref{bhpp}, we calculate the pair production rate for all instantons compatible with the given spacetime metric and in Section \ref{concl} we draw the conclusions.

\section{The noncommutative inspired Schwarzschild-deSitter spacetime \label{line}}

\subsection{The line element}

Before determining a new solution of Einstein equations, we need to recall the general ideas behind  noncommutative geometry inspired solutions. 
Being a model of quantum geometry, a noncommutative manifold undergoes  strong quantum fluctuations in the high energy/short distance regime. So before talking about length, line elements and other more sophisticated geometrical objects we need to understand the ultimate fate of a point and of all physical objects we are used to considering as point like.
A simple way to address this issue is to estimate the mean position of an object by averaging coordinate operators on suitable coordinate coherent states. As a result one finds that the mean position of a point like object in a noncommutative manifold is no longer governed by a Dirac delta function but by a Gaussian distribution
\begin{equation}
f_{\ell}(\vec{x})=\frac{1}{(4\pi\ell^2)^{d/2}} e^{-|\vec{x}|^2/4\ell^2},
\end{equation}
where $d$ is the manifold dimension and $\ell$, is the minimal length implemented through the noncommutative relations among coordinate operators \cite{NCQFT}. The value of $\ell$ is not fixed \textit{a priori}.  However, in the absence of extradimensions, a natural choice for $\ell$ would be a value of the order of the Planck length, i.e., $\ell\sim\sqrt{G}\approx 1.6\times 10^{-33}$ cm.
It has been shown that primary corrections to any field equation in the presence of a noncommutative background can be obtained by replacing the conventional point like source term (matter sector) with a Gaussian distribution, while keeping formally unchanged differential operators (geometry sector) \cite{review}. In the specific case of the gravity field equations this is equivalent to saying that the only modification occurs at the level of the energy-momentum tensor,  while $G_{\mu\nu}$ is formally left
unchanged.

For a static, spherically symmetric, noncommutative diffused, particle-like gravitational source of mass $M$, one gets a Gaussian profile for the $T_0^0$ component of the  energy-momentum tensor
 \begin{equation}
 T_0^0=-\rho_\ell (r)=-\dfrac{M}{\left(4\pi\ell^2\right)^{3/2}}\,\exp\left(-\frac{r^2}{4\ell^2}\right).
\end{equation}
The covariant conservation law $\nabla_\mu T^{\mu\nu}=0$ and the ``Schwarzschild like'' condition $g_{00}=-g_{rr}^{-1}$ completely specify the energy momentum tensor, whose form is given by
\begin{equation}
 T^\mu{}_\nu=\mathrm{Diag}\left(\, -\rho_\ell\left(\, r\,\right),\ p_r\left(\,r\,\right),\
 p_\perp\left(\, r\,\right),\ p_\perp\left(\, r\,\right)\,\right).
 \label{stresst}
 \end{equation}
We notice that there are nonvanishing pressure terms with $p_r\neq p_\perp$, corresponding to the case of an anisotropic fluid. Contrary to the conventional picture of matter squeezed at the origin,  here the noncommutative geometry
is effectively described as a fluid diffused around the origin. If one substitutes the above energy momentum tensor in Einstein equations one obtains the noncommutative geometry inspired Schwarzschild solution (NCSchw) \cite{ncschw}. At short distances the NCSchw solution is regular due to the noncommutative smearing effect. At large distances, namely for $r\gg \ell$, the above energy momentum tensor exponentially vanishes and one recovers the conventional vacuum solution, i.e., the Schwarzschild spacetime.

In this paper we want to determine black hole solutions in the presence of a background cosmological term. As a result we consider the Einstein equations
\be
R_{\mu\nu}-\frac{1}{2}Rg_{\mu\nu}+\Lambda g_{\mu\nu}=8\pi G T_{\mu\nu}
\ee
and a line element of the form
\be
 &&ds^2 = -V(r)\, dt^2 + V(r)^{-1}\, dr^2 +
r^2\,d\Omega^2. 
\ee
To solve Einstein equations it is convenient to introduce the following tensor
\be
&& {\cal T}_{\mu}^{\nu}\equiv T_{\mu}^{\nu}-\frac{\Lambda}{8\pi G}\delta_{\mu}^{\nu}=\\ &&= \mathrm{Diag}\left(\, {\cal E} \left(\,r\,\right),\ {\cal P}_r\left(\,r\,\right),\
 {\cal P}_\perp\left(\, r\,\right),\ {\cal P}_\perp\left(\, r\,\right)\,\right) \nonumber
\ee
with ${\cal E}(r)=-\rho-\Lambda/8\pi G$,
which yields the following ``fluid'' equations 
\begin{eqnarray}
&&\frac{d{\cal M}}{dr}= 4\pi\, r^2{\cal E}(r)\ ,\label{eq1}\\
&&\frac{1}{2g_{00}}\frac{dg_{00}}{dr}=G\ \frac{{\cal M}(r)+4\pi \ r^3 {\cal P}_r(r)}{r(r-2G{\cal M}(r))}\ , \label{eq22}\\
&&\frac{d{\cal P}_r}{dr}= - \frac{1}{2g_{00}}\frac{dg_{00}}{dr}\left(\, {\cal E}+{\cal P}_r\,\right)+\frac{2}{r}\left(\, {\cal P}_\perp-{\cal P}_r \,\right)
\label{eq33}
\end{eqnarray}
We recall that the condition $g_{00}=-g_{rr}^{-1}=-V(r)$ is equivalent to the equation of state
\begin{equation}
 {\cal P}_r(r)=-{\cal E}(r).
\end{equation}
As a result we obtain the NCSchwdS line element 
\be
V(r)= 1- \frac{4M G \gamma(3/2; r^2/4\ell^2)}{r\sqrt{\pi}}-\frac{\Lambda r^2}{3}\, \, 
\label{lineel}
\ee
where
\be
\gamma(3/2; x)=\int_0^x dtt^{1/2}e^{-t}.
\ee
The angular pressure turns out to be
\be
p_\perp (r)=-\rho_\ell(r)\left(1-\frac{r^2}{4\ell^2}\right).
\label{pperp}
\ee
We start the analysis of (\ref{lineel}), by noticing that for $r\gg \ell$ the solution coincides with the conventional Schwarzschild-deSitter line element. In other words this is the regime where noncommutative fluctuations are negligible and the spacetime can well described by a smooth differential manifold.
On the other hand, at small length scales, i.e., high energies there is a crucial departure from the conventional scenario. Expanding ($\ref{lineel}$) for $r\ll\ell$ we get
\begin{equation}
V(r)\approx 1-\frac{\Lambda_{\mathrm{eff}}}{3} r^2
\end{equation}
where 
\begin{equation}
\Lambda_{\mathrm{eff}}=\Lambda+\frac{1}{\sqrt{\pi}}\frac{MG}{\ell^3}.
\end{equation}
The metric is regular at the origin and we find a local deSitter spacetime whose cosmological constant $\Lambda_{\mathrm{eff}}$ is due to both the background cosmological term $\Lambda$ and the noncommutative fluctuations $\sim MG/\ell^3$.

\begin{figure}
 \begin{center}
 \includegraphics[width=7cm]{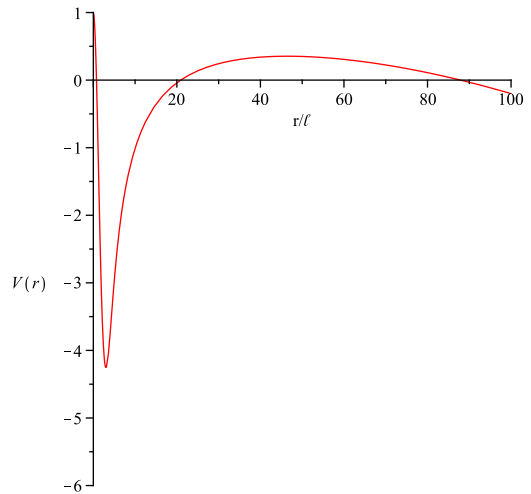}

 \hspace{0.2cm}
      \caption{\label{F1}The function $V(r)$ vs $r/\ell$ for $M=10\ell/G$ and $\Lambda/3=10^{-4}\ell^{-2}$. 
 }
 \end{center}
  \end{figure}

\begin{figure}
 \begin{center}
 \includegraphics[width=7cm]{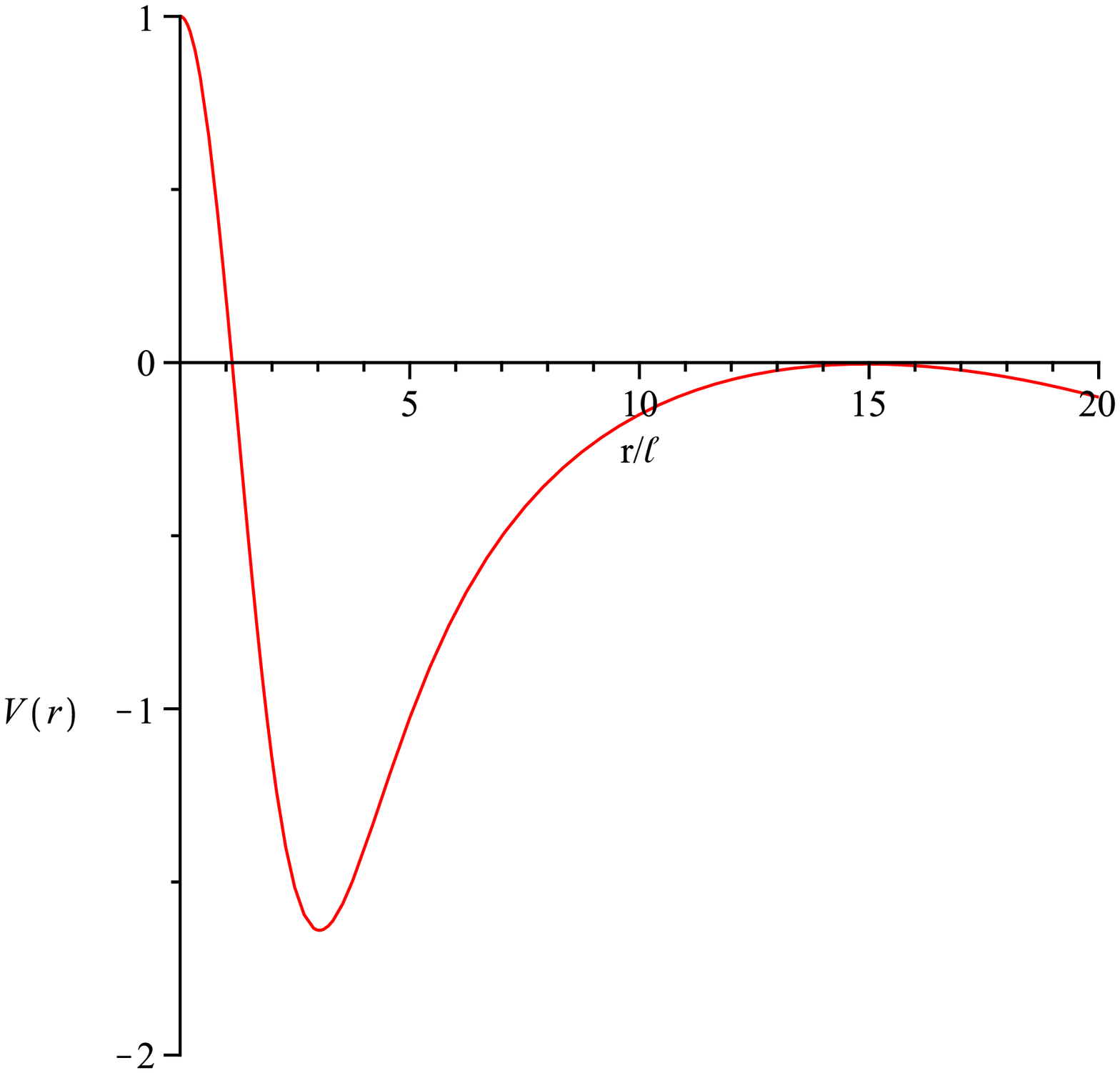}

 \hspace{0.2cm}
      \caption{\label{F7}The function $V(r)$ vs $r/\ell$ for $M=5\ell/G$ and $\Lambda/3=15\times 10^{-4}\ell^{-2}$. 
 }
 \end{center}
  \end{figure}

\begin{figure}
 \begin{center}
 \includegraphics[width=7cm]{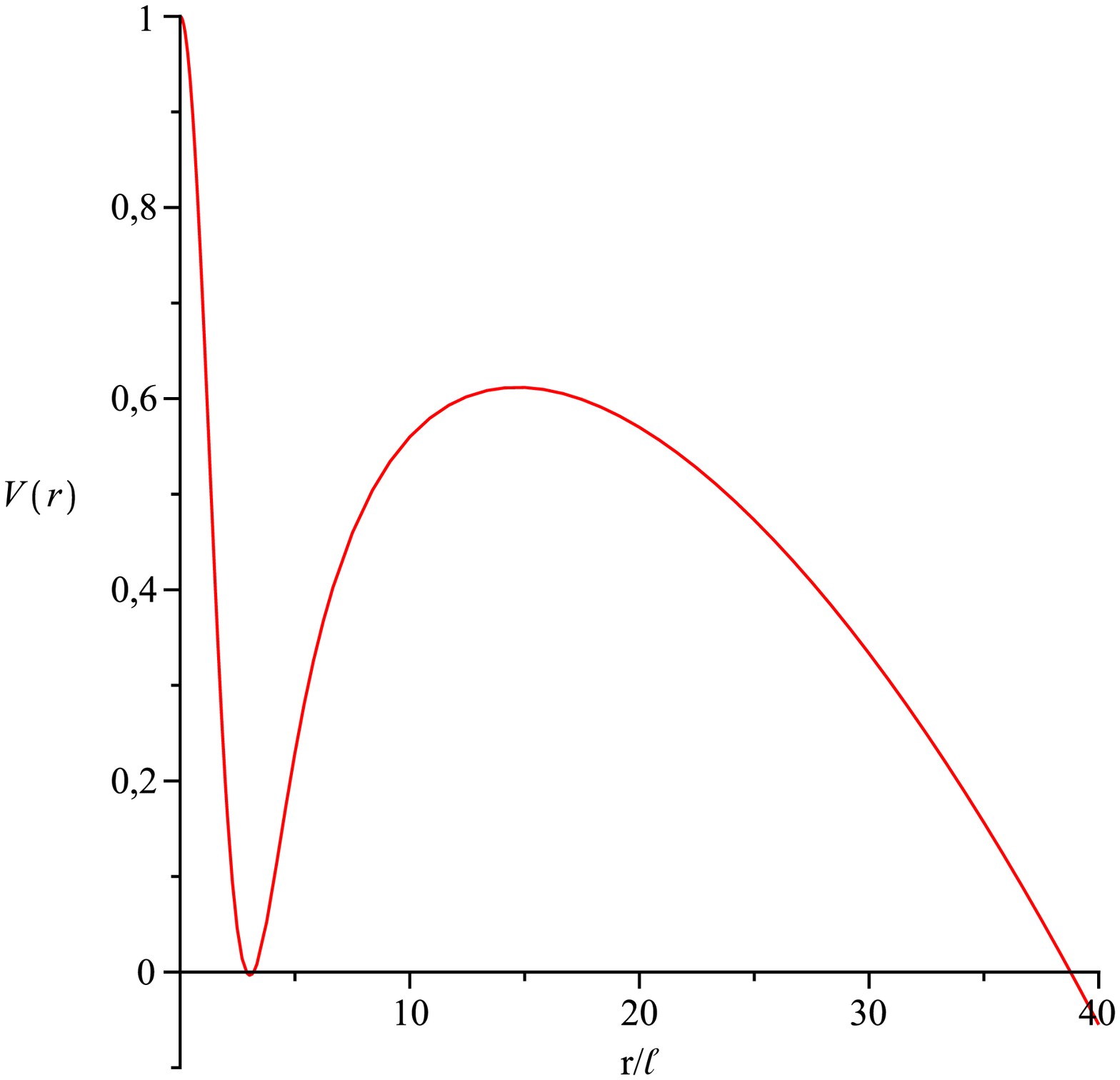}

 \hspace{0.2cm}
      \caption{\label{F5}The function $V(r)$ vs $r/\ell$ for $M=1.9\ell$ and $\Lambda/3=6\times 10^{-4}\ell^{-2}$.
 }
 \end{center}
  \end{figure}

\begin{figure}
 \begin{center}
 \includegraphics[width=7cm]{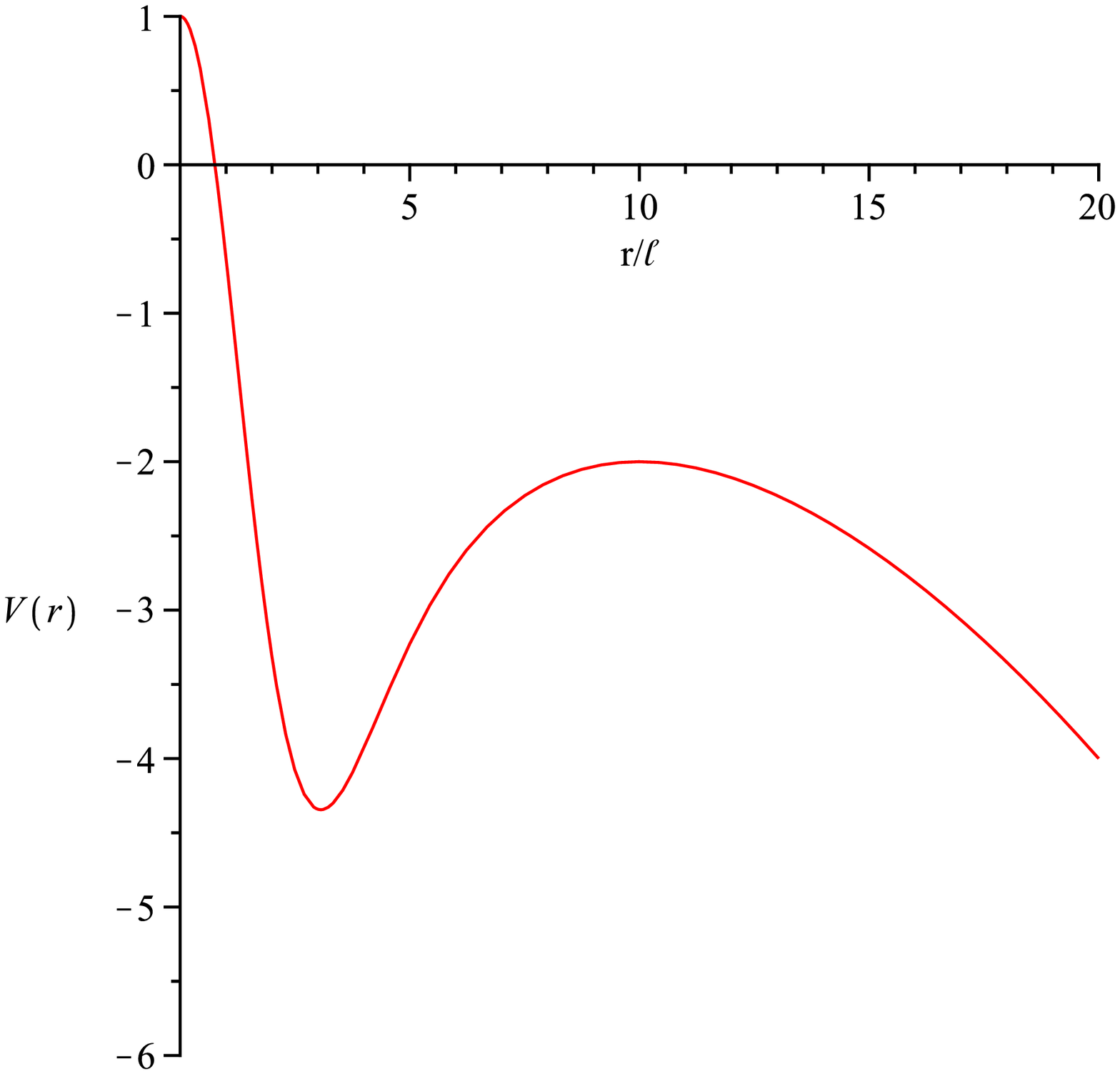}

 \hspace{0.2cm}
      \caption{\label{F2}The function $V(r)$ vs $r/\ell$ for $M=10\ell/G$ and $\Lambda/3=10^{-2}\ell^{-2}$.
 }
 \end{center}
  \end{figure}

\begin{figure}
 \begin{center}
 \includegraphics[width=7cm]{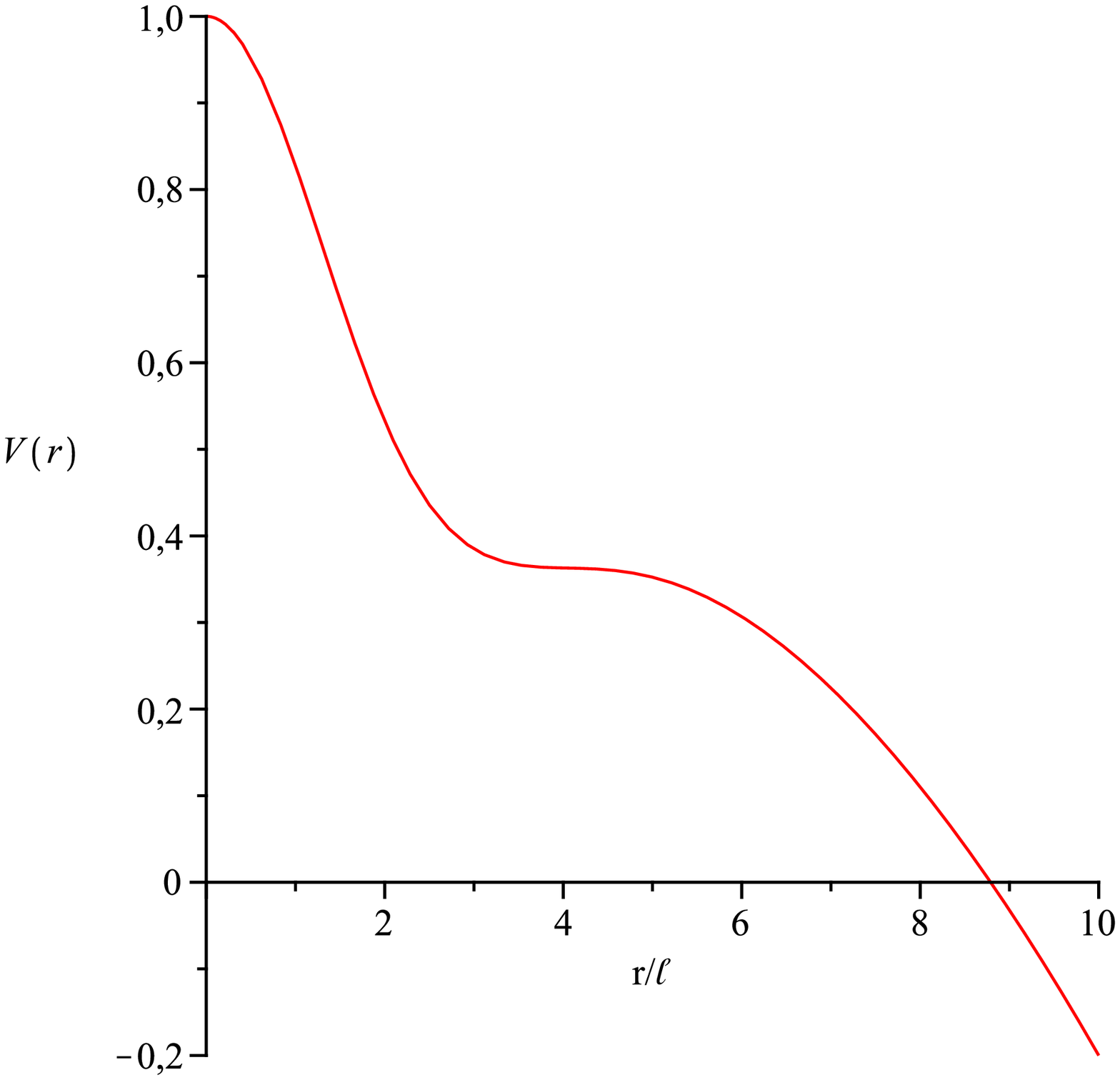}

 \hspace{0.2cm}
      \caption{\label{F3}The function $V(r)$ vs $r/\ell$ for $M=1\ell/G$ and $\Lambda/3=10^{-2}\ell^{-2}$.
 }
 \end{center}
  \end{figure}

\begin{figure}
 \begin{center}
 \includegraphics[width=7cm]{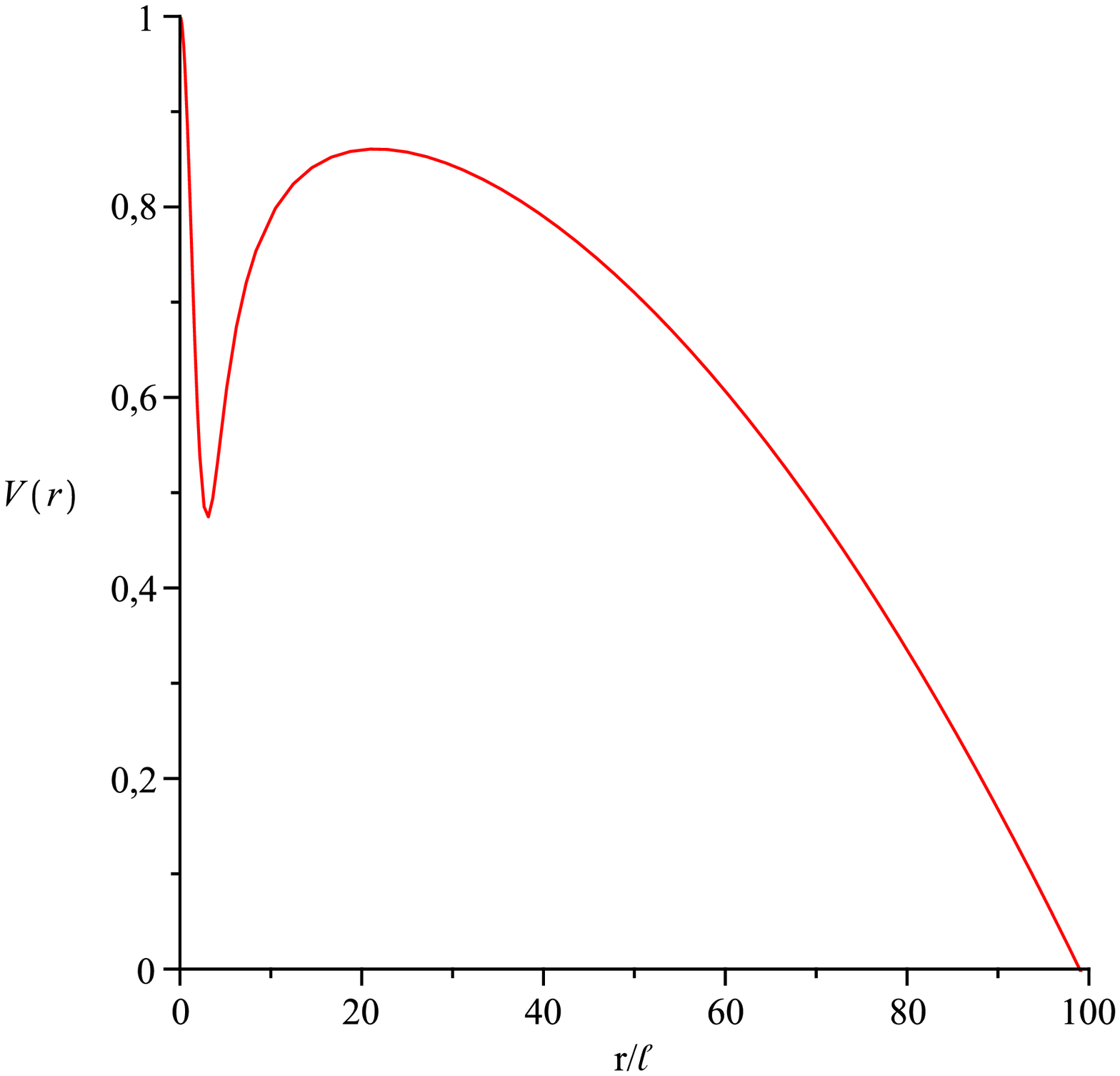}

 \hspace{0.2cm}
      \caption{\label{F4}The function $V(r)$ vs $r/\ell$ for $M=1\ell$ and $\Lambda/3=10^{-4}\ell^{-2}$.
 }
 \end{center}
  \end{figure}

An interesting feature of the solution is the horizon equation $V(r_H)=0$. This depends on two parameters, $M$ and $\Lambda$. As a result there is a mass $M_0=M_0(\Lambda)$ depending on $\Lambda$ such that
\begin{enumerate}
\item[a)\label{one}] for $M>M_0$ there are three horizons, an inner $r_-$ and an outer  black hole horizon $r_+$  and a cosmological horizon $r_c$ (see Fig. \ref{F1}).
\item[b)] \label{four}for $M=M_N>M_0$, the black hole outer horizon $r_+$ and the cosmological horizon $r_c$ coalesce into a single degenerate horizon $r_N$, corresponding to the case of a Nariai-like solution (see Fig. \ref{F7}).
\item[c)] \label{three}for $M=M_0$ the two black hole horizons coalesce into a single degenerate horizon $r_0$ and there is also a cosmological horizon $r_c$ (see Fig. \ref{F5}).
\item[d)] \label{two}$M<M_0$ there is just one  (cosmological) horizon (see Figs \ref{F2}, \ref{F3} and \ref{F4}), yielding a
soliton. 
\end{enumerate}
We recall that the NCSchw solution admits two, one or no horizon depending of the value of the mass parameter $M$. A difference between the two solutions is that $M_0$ now depends on the cosmological term. The cases a) and c) are  good approximations of the NCSchw black hole with two and one horizons respectively. The case b) is a novelty since for the NCSchw solution no Nariai configuration occurs.
Finally the case d) occurs in a variety of situations,  by decreasing the mass or increasing the cosmological term. For instance 
in Fig. \ref{F2} the internal horizon becomes the unique horizon and around $r\sim 10\ell$ there is a compensation of the massive and the cosmological term for negative values of $V(r)$, preventing the formation of any other horizon. This case has no analogue in the asymptotically flat space.
For lighter masses, the cosmological term dominates even at short distances and the horizon takes place at larger distances. The cases in Fig. \ref{F3} and in \ref{F4} are the analogues of the mini-gravastar case of the NCSchw solution, namely a regular geometry without black hole horizons. However  topologically the geometries in Fig. \ref{F2}, \ref{F3} and \ref{F4} are equivalent. 

  
\subsection{Negative mass solutions}

  \begin{figure}
 \begin{center}
 \includegraphics[width=7cm]{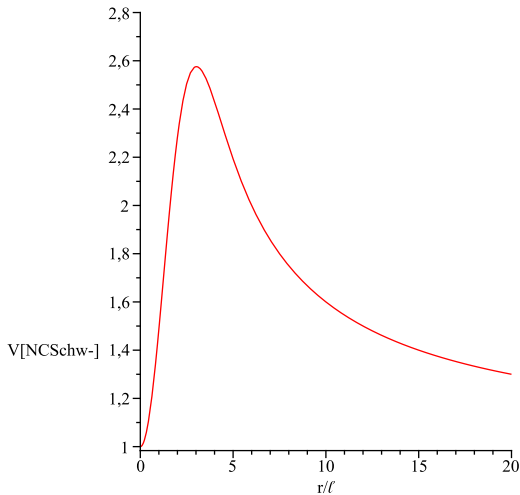}

 \hspace{0.2cm}
      \caption{\label{F8}The function $V_{\mathrm {NCSchw}-}(r)$ vs $r/\ell$ for $M=-3\ell$, corresponding to the negative mass noncommutative geometry inspired Schwarzschild solution.
 }
 \end{center}
  \end{figure}
  
    \begin{figure}
 \begin{center}
 \includegraphics[width=7cm]{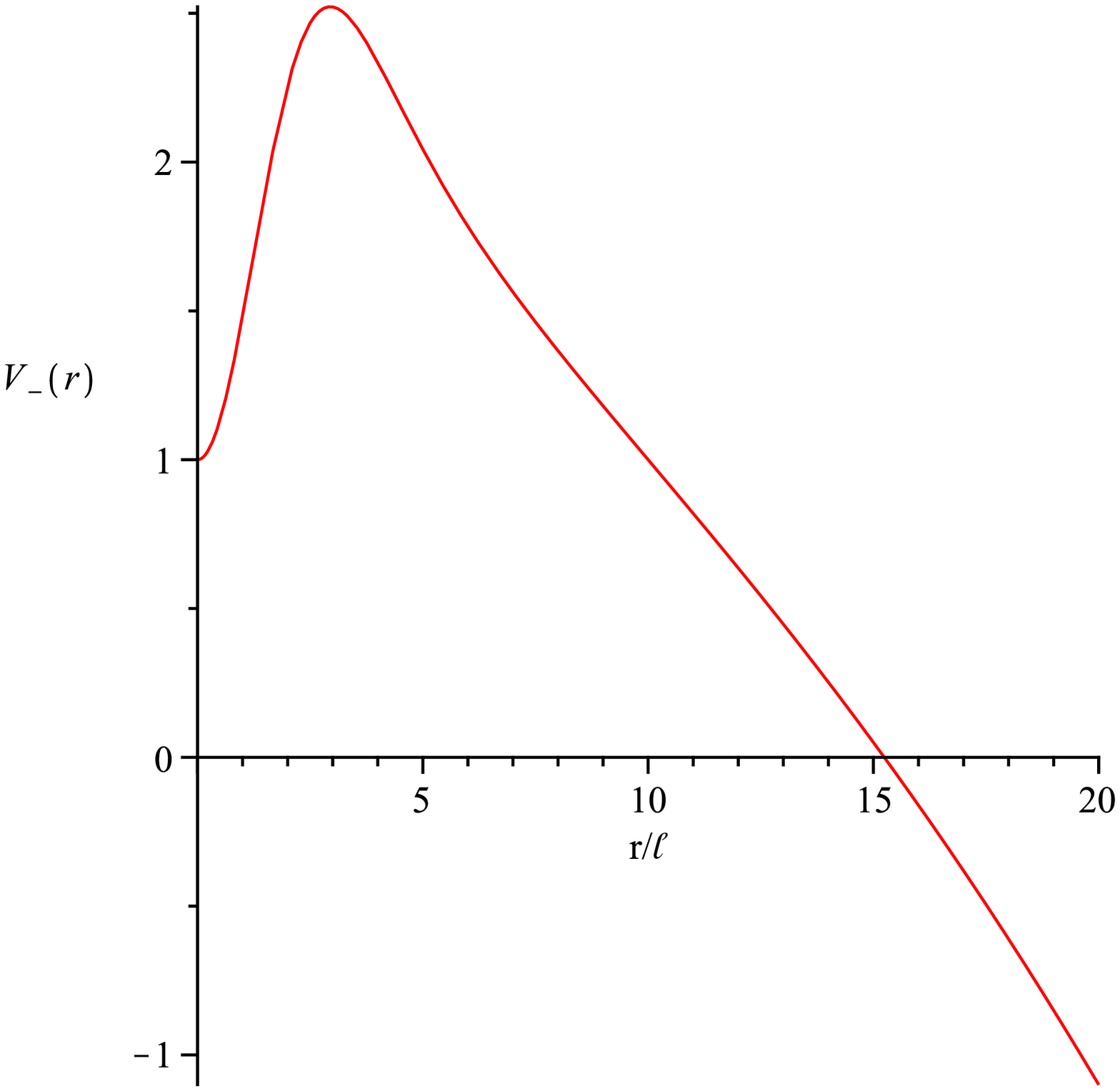}

 \hspace{0.2cm}
      \caption{\label{F9}The function $V_-(r)$ vs $r/\ell$ for $M=-3\ell$ and $\Lambda/3=6\times 10^{-3}\ell^{-2}$.
 }
 \end{center}
  \end{figure}

In view of the study of the stability of the deSitter background one must consider contributions coming from instantons, corresponding to all possible spacetime configurations.
To this purpose we must clarify the scenario when the radial coordinate assumes negative values. At first sight one may be tempted to say that we are continuing our spacetime through the origin. However we stress that the region of negative radial coordinate is not a continuation of the spacetime for positive $r$. Indeed the spacetime at $r=0$ is locally flat, and so the spacetime for positive $r$ is geodesically complete. This means that  we have two distinct spacetimes, one for $r>0$ and one for $r<0$, and both must be considered in the contribution to the action (\ref{piaction}). An equivalent and maybe more correct way to see this fact is to consider two distinct spacetimes having positive $M>0$ or either negative $M<0$ mass parameter. Regions of negative density might occur in spacetime as a result of quantum fluctuations of the vacuum energy density at sufficiently short distances and/or early times \cite{Morris:1988tu}. To this purpose it has been shown that negative energy density can undergo gravitational collapse to form a black hole \cite{Mann:1997jb}.  As a result we have an additional line element  with deSitter background for $M=-|M|$
\begin{equation}
V_-(r)= 1+\frac{4 |M| G \gamma(3/2; r^2/4\ell^2)}{r\sqrt{\pi}}-\frac{\Lambda r^2}{3}.
\end{equation} 
In the asymptotically flat case, $\Lambda=0$, we have the special case of the noncommutative geometry inspired Schwarzschild spacetime
\begin{equation}
V_{\mathrm {NCSchw}-}(r)= 1+\frac{4 |M| G \gamma(3/2; r^2/4\ell^2)}{r\sqrt{\pi}} 
\end{equation} 
 describing a soliton of  negative mass.
 
To discover the properties of these spacetimes, we start from the latter solution. In Fig. \ref{F8} we see that no horizon occurs. In addition the spacetime is regular and geodesically complete, being asymptotically flat both at the origin and infinity. Recall that this solution is generated by an energy momentum tensor of an anisotropic fluid as in (\ref{stresst}). However contrary to the positive mass solution for which negative pressures prevent the collapse of the energy density into a Dirac delta function, in the negative mass case we have a negative energy density whose expansion is contained by positive (inwards) pressure terms. 
The case with $\Lambda\neq 0$ is shown in Fig. \ref{F9}. We see that there is only a cosmological horizon.

\subsection{Thermodynamics}

  \begin{figure}
 \begin{center}
 \includegraphics[width=7cm]{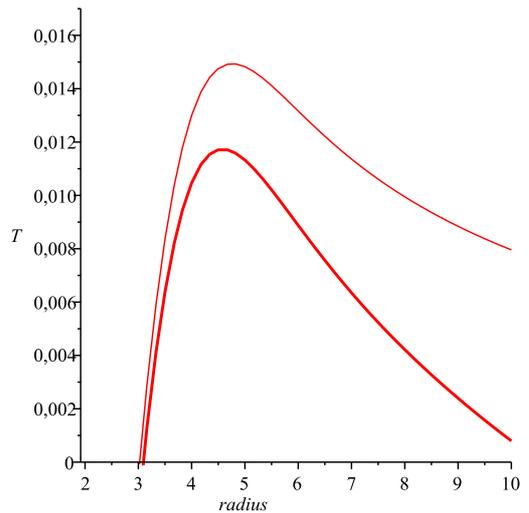}

 \hspace{0.2cm}
      \caption{\label{F6}The thin line represents the black hole temperature in a deSitter background, $T_+\times \ell$ as a function of $r_+/\ell$ for $\Lambda/3=3\times 10^{-3}\ell^{-2}$. The thick line represents the temperature of the NCSchwBH in asymptotically flat space. }
 \end{center}
  \end{figure}
  
To discover the thermodynamic properties of the solution let us start by considering 
the internal energy of the black hole $U(r_H)$. Following an approach analogous to that for the conventional Schwarzschild deSitter case, this is simply the mass parameter $M$ which becomes a function of the horizon by requiring $V(r_H)=0$.   Thus
\begin{equation}
M=U(r_H)=\frac{r_H \Gamma(3/2)}{2 G\, \gamma(3/2; r_H^2/4\ell^2)}\left(1-\frac{\Lambda}{3}r_H^2\right)
\label{massh}
\end{equation}
where $\Gamma(3/2)=\sqrt{\pi}/2$. For  convenience we introduce the function
$g(r)=\gamma(3/2; r^2/4\ell^2)/\Gamma(3/2)$. 
As a consequence 
\be
&&\frac{\partial U}{\partial r_H}=\\
&&\frac{1}{2 G\, g( r_H)}\left\{\left(1-r_H\frac{{g}^\prime (r_H)}{{g}(r_H)}\right)\left(1-\frac{\Lambda}{3}r_H^2\right)-\frac{2}{3}\Lambda r_H^2\right\}.\nonumber
\ee
By definition the temperature reads 
\be 
T_H=\frac{1}{4\pi}\left. \frac{dV(r)}{dr} \right|_{r_H}
\ee
where from (\ref{lineel})
\be 
\frac{dV(r)}{dr}=\frac{2G M g(r)}{r^2}\left(1-r\frac{g'(r)}{g(r)}\right)-\frac{2}{3}\Lambda r.
 \ee

As a consequence the temperature is
\begin{equation}
T_H=\frac{1}{4\pi r_H}\left\{ \left(1-r_H\frac{g^\prime (r_H)}{g(r_H)}\right)\left(1-\frac{\Lambda}{3}r_H^2\right)-\frac{2}{3}\Lambda r_H^2\right\}.
\label{temp}
\end{equation}
The above result holds also in the case of the negative mass solution $V_-(r)$.

Some comments are in order. Insofar as we have three horizons as in Fig \ref{F1}, the formula in (\ref{temp}) describes for $r_H=r_+$ the black hole temperature, where for convenience we write
\begin{eqnarray}
T_+&=&T_{\star}(r_+)-T_{\mathrm{dS}}(r_+)
\label{bhtemp}
\end{eqnarray}
where
\begin{equation}
T_{\star}(r_+)=\frac{1}{4\pi r_+} \left(1-\frac{r_+^3}{4\ell^3}\frac{e^{-r_+^2/4\ell^2}}{\gamma(3/2; r_+^2/4\ell^2)}\right)\left(1-\frac{\Lambda}{3}r_+^2\right)
\label{htemp}
\end{equation} 
approaches, in the regime $r_+\sqrt{\Lambda}\ll 1$, the black hole temperature in the asymptotically flat case  and
\begin{equation}
T_{\mathrm{dS}}(r_+)=\frac{\Lambda}{6\pi} \ r_+ 
\label{dstemp}
\end{equation}
is the deSitter background temperature at the black hole event horizon.
The background temperature  $T_{\mathrm{dS}}$ and the term depending on $\Lambda$ in $T_\star$  decrease the temperature $T_+$ with respect the corresponding value in asymptotically flat space (see Fig. \ref{F6}).

Black hole evaporation takes place in two phases. The first is the Hawking phase, where the temperature follows the law $1/r_H$ and the black hole heat capacity is negative. This continues until the temperature reaches a maximum, after which the second phase takes place.  This phase is  the cooling down phase and is characterized by a positive heat capacity. The latter phase takes place until the black hole shrinks to the radius $r_0$,  corresponding to the case of thermal equilibrium between the temperature $T_\star$ and the deSitter temperature $T_{\mathrm{dS}}$. Also the evaporation endpoint is modified with respect to the asymptotically flat case. At $r_0$ we have an extremal black hole remnant whose size is in general bigger than the corresponding value for the asymptotically flat case $\approx 3.0\ell$ (Fig. \ref{F6}).  On the other hand the remnant mass is in general lighter than the corresponding value for the asymptotically flat case due to the presence of the deSitter term in (\ref{massh}). However, in the limit $r_+\sqrt{\Lambda}\ll 1$, the deSitter terms are negligible and one recovers the usual temperature for the asymptotically flat NCSchw solution
\begin{equation}
T_+\approx T_\star.
\end{equation}

For $r_H=r_c\sim 1/\sqrt{\Lambda}\gg r_+$ the formula in (\ref{temp}) no longer describes the temperature of the black hole but rather that of the cosmological horizon. To get the correct expression for the temperature one has to consider the absolute value of (\ref{temp}) to obtain a positive defined quantity. As a result one finds  
\begin{eqnarray}
T_c&=&T_{\mathrm{dS}}(r_c)+T_\star(r_c)
\label{ctemp}
\end{eqnarray}
which is the the conventional deSitter bath temperature $T_{\mathrm{dS}}(r_c)$ plus the black hole temperature at $r_c$. 
For $r_c\gg \ell$ the dominant correction coming from $T_\star(r_c)$ is
\begin{equation}
T_\star(r_c)\approx\frac{\Lambda}{12\pi}r_c.
\end{equation}
This tells us that the black hole  thermalizes the deSitter background, which reaches a temperature higher than the conventional case in the absence of black holes. As a result we find 
\begin{equation}
T_c\approx  \frac{\Lambda}{4\pi} \ r_c.
\end{equation}
which exceeds $T_{\mathrm{dS}}(r_c)$.
For $r_c\gg MG$, the cosmological horizon approaches the conventional value $r_c\approx\sqrt{3/\Lambda}$. Therefore the temperature reads
\begin{equation}
T_c\approx  \frac{3}{4\pi}\sqrt{\frac{\Lambda}{3}}.
\end{equation}

In the case of a single horizon, (\ref{temp}) still holds and one can determine the temperature for gravitational objects described in Fig. \ref{F2}, \ref{F3} and \ref{F4}. We shall get back to these cases in the section about the gravitational instantons.
\section{The anisotropic fluid functional action \label{faction}}


The production of black holes in the deSitter background is governed by the Euclidean version of the action 
 \be
I=\int d^4 x\sqrt{-g}\left[\frac{1}{16\pi G}\left(R-2\Lambda \right)+L_m\right]
\label{action}
 \ee
generating gravitational field equations.
To this purpose, we still need to find the form of the Lagrangian $L_m$, supposed to lead to the noncommutative Schwarzschild solution. We will proceeds along the line of Ref. \cite{Brown:1992kc}. 
Our effective fluid-like approach 
takes leading order quantum geometry effects into account while  
letting us formally work in a classical framework. 
To begin, we recall the basic  notation  
 for the case of an isotropic fluid. Then we will extend the formalism to the anisotropic fluid case. 

 An isotropic fluid has a generic energy density $\rho$ and a unique pressure $p_r=p_\perp\equiv p$.  In addition to these variables, we can introduce the standard notation for the spacetime scalar fields
  \be
 n&=&\mathrm{particle\ number\ density}\\
  T&=&\mathrm{temperature}\\
  s&=&\mathrm{entropy \ per\ particle}
  \ee
  whose values represent measurements made in the rest frame of the fluid. The fluid motion can be characterized  by its unit four velocity vector field $u^\mu$, which in the rest frame reads
  \be
  u^{\mu}=\frac{1}{\sqrt{-g_{00}}}\left(-1,0,0,0\right).
  \ee
  It is also convenient to define 
  \be
  \mu &\equiv &\frac{\rho+ p}{n}=\mathrm{chemical \ potential.}
  \ee
  The chemical potential is the energy per particle required to inject a small amount of fluid into a fluid  sample, keeping the sample volume and the entropy per particle $s$ constant. Of course in case of anisotropicity, chemical potentials depend on the direction of the injection due to the occurrence of different pressures. 
 The above thermodynamic variables are related by the local expression of the first law of thermodynamics, namely
  \be 
  d\rho=\mu dn+nT\ ds
  \ee
 showing that the equation of state for the fluid can be specified by giving the function $\rho(n, s)$, the energy density as a function of the number density and the entropy per particle.
  The equations of motion of a perfect fluid, both isotropic or anisotropic, consist of stress-energy conservation, namely $\left. T^{\mu\nu}\right. _{;\nu}=0$, and the equation 
 $
 \left(nu^{\mu}\right)_{;\mu} = 0
 $
  expressing conservation of particle number.  
  The action functional we are going to present here provides the stress tensor
  \be
  T^{\mu\nu}=\rho u^\mu u^\nu + p\left( g^{\mu\nu}+u^\mu u^\nu\right),
  \ee  
  its conservation and the first law of thermodynamics, given the equation of state $\rho (n, s)$. The fluid action $S$ is a function of $j^\mu\equiv \sqrt{-g} n u^\mu$. As a result the fluid four velocity can be written as
  \be
  u^\mu=j^\mu/|j|
  \ee
  where $|j|=\sqrt{-j^\mu g_{\mu\nu}j^\nu}$. The particle number density is therefore given by 
  \be
  n=|j|/\sqrt{-g}.
  \ee

As explained in \cite{Brown:1992kc}, for our specific purposes we just need to consider the ``on shell'' action, namely
 \be
 S(\mathrm{on\ shell})=\int d^4 x \sqrt{-g}\ p (n, s).
 \ee
with the equation of state $\rho=-p$.


The above analysis can be extended to the anisotropic fluid case in order to generate the NCSchwdS geometry. To this purpose, we follow the results in \cite{psl} to determine the ``on shell'' action. It is convenient to introduce a new vector $k^\mu\equiv \tilde{n}\sqrt{-g} \ l^\mu$, where 
\be
l^{\mu}=\frac{1}{\sqrt{g_{rr}}}\left(0,1,0,0\right).
\ee
and $\tilde{n}$ is a generic particle number density to be identified later. As a result we have
\be
\tilde{n}=|k|/\sqrt{-g},
\ee
where $|k|=\sqrt{g_{\mu\nu}k^\mu k^\nu}$.
We begin by writing the action $I_m(\mathrm{on\ shell})=\int d^4 x\sqrt{-g} \ L_m$ as
\be
&& I_m(\mathrm{on\ shell})=\\&& =\int d^4 x\sqrt{-g} \ p_r\left( n, s\right)+\int d^4 x\sqrt{-g}\ L_\perp\left( n,\tilde{n}, s\right)\nonumber
\ee
where the first term in the square brackets gives the isotropic fluid energy momentum tensor 
\be
T^{\mu\nu}_{\mathrm{iso}}=\rho_{\ell}u^\mu u^\nu + p_r\left( g^{\mu\nu}+u^\mu u^\nu\right),
\ee 
with equation of state $p_r=-\rho_{\ell}$, while $L_\perp$ provides the breaking of the isotropy. As a consequence the energy momentum momentum tensor whose action we are looking for is nothing but
 \be
T^{\mu\nu}=T^{\mu\nu}_{\mathrm{iso}}+\left(p_\perp -p_r\right)\left( g^{\mu\nu}+u^\mu u^\nu -l^\mu l^\nu \right).
\label{aniso}
\ee 
namely an isotropic term plus corrections coming from variations of $L_\perp$. The latter can be computed and read
\be
T^{\mu\nu}_{\perp}&=&\left(L_\perp-n\frac{\partial L_\perp}{\partial n}-\tilde{n}\frac{\partial L_\perp}{\partial \tilde{n}}\right)g^{\mu\nu}+\\&&+ \tilde{n}\frac{\partial L_\perp}{\partial \tilde{n}}\ l^\mu l^\nu -n\frac{\partial L_\perp}{\partial n}\ u^\mu u^\nu . \nonumber
\ee
If we want to reproduce the form of (\ref{aniso}), we have that
\be
L_\perp=\tilde{n}\frac{\partial L_\perp}{\partial \tilde{n}}= {n}\frac{\partial L_\perp}{\partial {n}}=-\left(p_\perp -p_r\right).
\ee
Thus without loss of generality we can assume $n=\tilde n$, with $n=n\left(|j|/\sqrt{-g}, |k|/\sqrt{-g}, s\right)$, and write the action as
\be
S(\mathrm{on\ shell})=\int d^4 x \sqrt{-g}\ \left(2p_r-p_\perp\right).
\ee
Employing Eq. (\ref{pperp}), one finds that
\be
L_m(\mathrm{on\ shell})&=&p_r + \frac{r^2}{4\ell^2}\frac{M}{\left(4\pi\ell^2\right)^{3/2}} e^{-\frac{r^2}{4\ell^2}}
\label{lagranis}
\ee
and finally
\be
T^{\mu\nu}=\left(\rho_{\ell}+p_\perp\right)\left(u^\mu u^\nu-l^\mu l^\nu\right)+p_\perp g^{\mu\nu}.
\ee
We notice that in (\ref{lagranis}) we have, in addition to the usual pressure term as in the isotropic case, another term which is due to the fact that $p_r\neq p_\perp$. 

\section{Pair creation rates 
\label{bhpp}}

\subsection{Gravitational instantons}

The pair production of objects in a cosmological background is described by propagation from nothing to a surface $\Sigma$, whose topology depends on the kind of instanton considered \cite{EQG}. 
The amplitude for these processes will be given by the path integral in (\ref{piaction}), over all metrics which agree with given boundary data $h_{ij}$ on $\Sigma$. 


In general a Euclidean solution will not match onto its real Lorentzian counterpart according to the standard 
$t\to i\tau$ prescription,  and it is not possible to require  that the instanton both be real and match its Lorentzian counterpart
along a $t =$constant hypersurface \cite{Ivan}, though for diagonal metrics both requirements can be satisfied. The matching conditions are the only conditions available that prescribe the connection between the instantons and the physical Lorentzian solutions, and so we match on a hypersurface $\Sigma$ whose
extrinsic curvature vanishes.  Consequently 
 $\Sigma$ can be interpreted as the zero momentum initial data for the Lorentzian extension of the solution \cite{Mann:1995vb}.  

By analytically continuing $t\to i\tau$ one gets the Euclidean line element
\be
 &&ds_E^2 = V(r)\, d\tau ^2 + V(r)^{-1}\, dr^2 +
r^2\,d\Omega^2 .
\ee
The spacetime is defined only for regions where the function $V(r)$ assumes positive values.
The corresponding Euclidean action is the Wick rotated version of (\ref{action}) and reads
\be
I_E&=&-\int_{\mathbb{M}_+} d^4 x\sqrt{g}\left[\frac{\Lambda}{8\pi G}-\frac{\mathsf{T}}{2}+L_m\right]\\
&+&(\mathrm{gravitational \ boundary \ terms})\nonumber
\ee
where $ \mathsf{T}=T^\mu_\mu$.   Here $\mathbb{M}_+$ is one of the parts the surface $\Sigma$ divides the (simply connected) spacetime $\mathbb{M}$ into. Since amplitudes due to each part are equal, we need only consider the path integral (\ref{piaction}) over all metrics on the half manifolds, e.g. $\mathbb{M}_+$.
The above Euclidean action describes the deSitter universe with nucleation of gravitational objects. For this reason we shall refer to it as $I_{\mathrm{obj}}$.
Note that since we identify $\Sigma$ with a surface of zero extrinsic curvature in the Euclidean section, the gravitational boundary term will not contribute to the action. On the other hand, if there is a Euclidean classical solution, i.e., an instanton which interpolates within the given boundary, the integral is dominated by the contribution coming from it.
There exists another relevant topology, which describes the deSitter background  universe without nucleation, i.e. $I_{\mathrm{bg}}$. We can determine this instanton from
\begin{equation}
I_{\mathrm{bg}}=-\int_{\mathbb{M}_+} d^4 x\sqrt{g}\ \frac{\Lambda}{8\pi G}=-\frac{3}{2}\frac{\pi}{\Lambda G}
\end{equation}
which gives the probability at which deSitter space is itself created
\begin{equation}
P_{\mathrm{bg}}=e^{\frac{3\pi}{\Lambda G}}.
\end{equation}
 As explained in the introduction the ratio of the two probability measures $P_{\mathrm{obj}}/P_{\mathrm{bg}}=e^{-2(I_{\mathrm{obj}}-I_{\mathrm{bg}})}$ gives the rate of  pair creation on an inflationary background, $\Gamma$ in (\ref{bhbgrate}).
The object created depends on the properties of the function $V(r)$. We recall that in the Lorentzian section $0\le r<\infty $ the spacetime is geodesically complete for positive as well negative mass parameter.
For positive mass $M>0$, the function $V(r)$ can have three roots, i.e., $r_-$, $r_+$ and $r_c$ (see Fig. \ref{F1}). To obtain a positive-definite metric, we must restrict $r$ to $r_+\le r\le r_c$. However the resulting instanton might be singular for a conical singularity at $r=r_+$ and $r=r_c$. In the degenerate case (see Fig. \ref{F5}), i.e. for $r_-=r_+$, the range of $r$ in the Euclidean section will be $r_+<r\le r_c$, as the double root in $V(r)$ implies that the proper distance from any other point to $r=r_+$ along spacelike directions is infinite. As a consequence we may obtain a regular instanton by identifying $\tau$ periodically with period $2\pi/\kappa_c$, where $\kappa_c$ is the surface gravity of the cosmological horizon. Following \cite{Romans:1991nq}, this instanton will be referred to as a \textit{cold instanton}. For non degenerate horizons, we must have the same period for $\tau$ and this corresponds to requiring 
\begin{equation}
\kappa_+=\kappa_c.
\label{sgc}
\end{equation}
As a special case we have the \textit{Nariai instanton} for which $r_+=r_c$. In general, i.e., for $r_+\neq r_c$, the condition (\ref{sgc}) is satisfied by a \textit{lukewarm instanton}, following the definition in \cite{Romans:1991nq}. 

In addition to these instantons, there are those corresponding to topologies in Figs \ref{F2}, \ref{F3}, \ref{F4}, which does not have any classical analogue. These cases occur when $V(r)$ has just a single root $r_1$. In this group of new topologies we need to consider also the negative mass topology which exhibits a single horizon (see Fig. \ref{F9}). This means that, as far as pair production is concerned, we always integrate the
action from $r=0$ outward to the single horizon $r_1$, i.e. $0\le r\le r_1$.   The nature of the single horizon will be equivalent to that of any single cosmological horizon, regardless of its size with respect the cosmological scale $1/\sqrt{\Lambda}$.

\subsection{Black hole pair production}

\begin{figure}
 \begin{center}
 \includegraphics[width=5.5cm,angle=0]{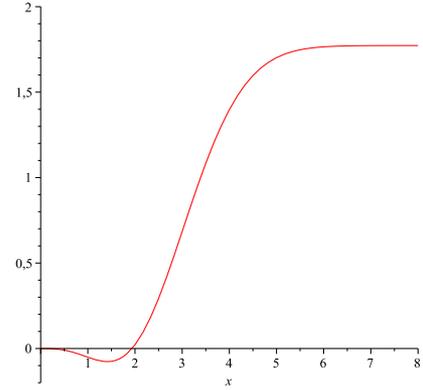}
 \hspace{0.2cm}
      \caption{\label{gammaminus} The function $2 \gamma(3/2;x^2/4) - \frac{1}{4}\ x^{3} e^{ - x^2/4}$ vs $x$. 
 }
 \end{center}
  \end{figure}

We now explicitly calculate black hole pair production rates.  For notational convenience we start by introducing 
\begin{equation}
 I_\Lambda=\int_{\mathbb{M}_+} d^4 x\sqrt{g}\ \frac{\Lambda}{8\pi G}
\end{equation} 
 and
\be
I_M=\frac{M}{(4\pi\ell^2)^{3/2}}\int_{\mathbb{M}_+} d^4 x\sqrt{g}\  \left(\frac{r^2}{2\ell^2}-1\right)e^{-r^2/4\ell^2}.
\ee
 Up to boundary terms, the Euclidean action $ I_{\mathrm{obj}}$ can be cast in the form $I_{\mathrm{obj}}=-I_M-I_\Lambda$.
In the lukewarm instanton case, we need to compute the integral
\[
 I_{\Lambda}=\int_{\mathbb{M}_+} d^4 x\sqrt{g}\ \frac{\Lambda}{8\pi G}=\frac{\Lambda\beta_c}{12 G}\left(r_c^3-r_+^3\right)
\] 
where $\beta_c=T_c^{-1}=2\pi/\kappa_c$.  If the cosmological horizon $r_c\gg r_+$, we have 
\begin{equation}
I_{\Lambda }\approx\frac{\pi}{\Lambda G}\left[1-\left(r_+\sqrt{\Lambda/3}\right)^3\right]
\end{equation}
for $r_c\approx\sqrt{3/\Lambda}$. Furthermore, if $r_+\sqrt{\Lambda}\ll 1$ we can consider just the first term, i.e., 
\begin{equation}
I_{\Lambda }\approx\frac{\pi}{\Lambda G}.
\end{equation}
We can now calculate 
\[
I_{M}=\frac{M}{(4\pi\ell^2)^{3/2}}\int_{\mathbb{M}_+} d^4 x\sqrt{g}\  \left(\frac{r^2}{2\ell^2}-1\right)e^{-r^2/4\ell^2}
\]
recalling that $r_+>r_0\ge 3.0\ell$.  As a result one finds
\be
I_{M}=\frac{M\beta_c}{\sqrt{\pi}}\left[2 \gamma(5/2;x^2/4) -\gamma(3/2;x^2/4)\right]_{r_+/\ell}^{r_c/\ell}.
\label{itwo}
\ee
Since
\be \gamma(5/2;x^2/4) = \frac{3}{2}\ \gamma(3/2;x^2/4) - \frac{1}{8}\ x^{3} e^{ - x^2/4}\nonumber  
\ee
we can express (\ref{itwo}) in terms of $\gamma(3/2;x^2/4)$ only 
\be
I_{M}=\frac{M\beta_c}{\sqrt{\pi}}\left[2 \gamma(3/2;x^2/4) - \frac{1}{4}\ x^{3} e^{ - x^2/4}\right]_{r_+/\ell}^{r_c/\ell} .
\ee
For large values of the argument, $x\gg 1$, we have 
\[ 2 \gamma(3/2;x^2/4) - \frac{1}{4}\ x^{3} e^{ - x^2/4}\approx 2 \gamma(3/2;x^2/4)\approx \sqrt{\pi},\]
while for small values of the argument, $x\ll 1$, we have
$ \gamma(3/2;x^2/4)\approx x^{3}/12 $, and 
\[ 2 \gamma(3/2;x^2/4) - \frac{1}{4}\ x^{3} e^{ - x^2/4}\approx -\frac{1}{12}\ x^{3} \]
a negative vanishing value. From Fig. \ref{gammaminus}, we see that $2 \gamma(3/2;x^2/4) - \frac{1}{4}\ x^{3} e^{ - x^2/4}$ is zero for $x\approx 2$  whereas it rapidly
asymptotes to  $\sqrt{\pi}$ for any $x>6$. 


As a result, for the cold instanton we have 
\begin{equation}
I_{\mathrm{cold}}\approx -\frac{\pi}{\Lambda G}\left(1+4M_0G\sqrt{\frac{\Lambda}{3}}\left[1-\frac{0.67}{\sqrt{\pi}}\right]-(r_0^2\Lambda/3)^{3/2}\right) ,
\end{equation}
being $r_0\approx 3.0\ell$, 
\be 
2 \gamma(3/2;r_0^2/4\ell^2) - \frac{1}{4}\ (r_0/\ell)^{3} e^{ - r_0^2/4\ell^2}\approx 0.67. 
\nonumber
\ee
 and assuming $r_c\gg \ell$. 
 To obtain the black hole remnant pair production rate we consider the probability measure 
$P_{\mathrm{cold}}=e^{-2I_{\mathrm{cold}}}$ and we divide this by the probability measure for a universe without black holes, i.e. $P_{\mathrm{bg}}$. This yields 
\begin{equation}
\Gamma_{\mathrm{cold}}\approx e^{-\frac{\pi}{\Lambda G}\left(1-8M_0G\sqrt{\frac{\Lambda}{3}}\left[1-\frac{0.67}{\sqrt{\pi}}\right]+2(r_0^2\Lambda/3)^{3/2}\right)} 
\end{equation}
and so the probability for pair creation of cold NCBHs is very low, unless $\Lambda$ is close to the Planck value $\Lambda G=1$.
However for large $\Lambda$ black holes do not occur, unless for masses $M\gg\ell/G$. Thus in place of Planck-sized black holes we have the production of single horizon spacetimes (see Fig. \ref{F2}, \ref{F3} and \ref{F4}). We shall discuss the nature of these gravitational objects in the next section. 
For the lukewarm instanton case, the contribution of $I_M$ is vanishing. However the term $(r_+^2\Lambda/3)^{3/2}$ coming from $I_\Lambda$ can grow with respect to the cold case. As a result we get 
\begin{equation}
\Gamma_{\mathrm{lw}}\approx e^{-\frac{\pi}{\Lambda G}\left(1+2(r_+^2\Lambda/3)^{3/2}\right)}.
\end{equation}
These extremely suppressed rates are in agreement with results found in \cite{Bousso:1996au} for the Schwarzschild deSitter spacetime.

\subsection{Nariai and other instantons}

To exclude instabilities of the deSitter background we need to conclude our analysis by studying the remaining topologies, namely the Nariai instanton and the single horizon geometries.
For the Nariai instanton, where $r_+=r_c$, we can make a coordinate transformation and rewrite the metric as
\begin{equation}
ds_E^2=\frac{1}{A} \left( d\chi ^2 +\sin^2 \chi d\psi ^2\right)+\frac{1}{B} \left( d\vartheta ^2 +\sin^2 \vartheta d\phi ^2\right) 
\end{equation}
where $A$ and $B$ are constant, $\chi$ and $\vartheta$ run from $0$ to $\pi$, and $\psi$ and $\phi$ are  periodic coordinates with period $2\pi$. The instanton has topology $S^2\times S^2$, the direct product of two spheres with different radii. The condition $r_+=r_c$ implies that the origin retreats to infinite proper distance and so there is no longer any global event horizon. The Lorentzian section is just the direct product of two dimensional deSitter space and a two sphere of fixed radius, $dS^2\times S^2$.  Consequently this instanton does not represent pair creation of black holes, though higher-order quantum corrections are expected to break the degeneracy of the two roots \cite{Mann:1995vb}, rendering this solution equivalent to that of the NCBH of the previous section.

Finally we are left with the case of a single cosmological horizon $r_1$, which occurs in Fig. \ref{F2}, \ref{F3} and \ref{F4} or for negative masses.
The instanton in this case is calculated for $0\le r\le r_1$ assuming $\tau=0, \beta_1/2$ with  $\beta_1=2\pi/\kappa_1$, i.e., 
\begin{eqnarray}
I_1=\frac{\beta_1 M}{\sqrt{\pi}}\ \left[2 \gamma(3/2;r_1^2/4\ell^2) - \frac{1}{4}\ \frac{r_1^3}{\ell^3} e^{ - r_1^2/4\ell^2}\right]-\frac{\beta_1 \Lambda }{12G} \ r_1^3. 
\nonumber
\end{eqnarray}
This leads to the following pair creation rate 
\begin{equation}\label{singlepair}
\Gamma_1=e^{\frac{\beta_1 \Lambda }{6G} r_1^3}\ e^{-\frac{2\beta_1 M}{\sqrt{\pi}} \left[2 \gamma(3/2;r_1^2/4\ell^2) - \frac{1}{4}\ \frac{r_1^3}{\ell^3} e^{ - r_1^2/4\ell^2}\right]}\ e^{-\frac{3\pi}{\Lambda G}} .
\end{equation}

We now demonstrate that production of such single-horizon objects is negligible, unless $\Lambda G \sim 1$.  For small
$r_1$, a non-negligible pair production might occur only if
\begin{equation}
\beta_1 r_1^3 \left(\frac{ \Lambda }{6G} + \frac{M}{6\sqrt{\pi}\ell^3}\right)  >\frac{3\pi}{\Lambda G}.
\label{ppc}
\end{equation}
For small $r_1$  the temperature increases (see Fig. \ref{F2}), then $\beta_1$ is little and this condition cannot be met. Indeed in the regime $r_1\sim\ell$ the temperature is
\begin{equation}
T_1\approx \frac{3}{4\pi r_1}
\end{equation}
and we have
\begin{equation}
\frac{4\pi r^4_1}{3}\left(\frac{ \Lambda }{6G} + \frac{M}{6\sqrt{\pi}\ell^3}\right) >\frac{3\pi}{\Lambda G}
\end{equation} 
Hence for $r_1\ll 1/\sqrt{\Lambda}$ the condition (\ref{ppc}) is not fulfilled for either small $M$ or large $M$ (as the latter 
will yield two horizons and thus lukewarm instantons) and no pair production occurs.
Indeed the pair production rate is
\begin{equation}
\Gamma_{1}\approx e^{-\frac{3\pi}{\Lambda G}}.
\end{equation}
The interpretation of this result is the following. The horizon in Fig. \ref{F2} is similar to a cosmological horizon,  apart from its size. The above result tells us that Planck size deSitter space does not occur in present times, in accord with experience. Conversely it could have been plentifully produced in early epochs of the universe, i.e, for $\Lambda G\sim 1$. Indeed this instanton occurs when the cosmological term $\Lambda$ overcomes the mass term $M$ in the NCSchwdS solution. This means that in place of Planck size black holes, this single horizon geometry would have been favoured during inflation. However, this instanton can be simply considered equivalent to the deSitter space when $\Lambda G\sim 1$ and cannot be considered as a source of instability.

For large horizon radii, the temperature $T_1\approx T_c$ is   very low and $\beta_1$ can significantly grow (see Fig. \ref{F3}, \ref{F4} and \ref{F9}). 
However we have
\begin{equation}
\frac{ \Lambda r_1^3}{6G T_c}\left( 1- \frac{12GM}{\Lambda r^3_1 }\right) \lesssim \frac{3\pi}{\Lambda G}
\end{equation}
since for single horizon objects we have  $\sqrt{\Lambda} r_1 \lesssim 1$ and $GM\sqrt{\Lambda}   \lesssim 1$. This is
in agreement to what found for the cold and lukewarm case. As a result the pair production rate is higher but yet largely suppressed unless for Planckian values of the cosmological constant, i.e.,
\begin{equation}
\Gamma_{\mathrm{1}}\approx e^{-\frac{\pi}{\Lambda G}}.
\end{equation}
This result corresponds to the fact that the lowest limit for the temperature, i.e., the highest value for $\beta_1$ is given by the deSitter universe in the absence of pair production $T_{dS}=(2\pi)^{-1}\sqrt{\Lambda/3}$.

 If $M<0$ then the inequality in (\ref{ppc}) can be satisfied, rendering the possibility that copious production of such objects
could perhaps be responsible for dark energy.  However  the production rate grows exponentially with increasing
$|M|$, leading to an instability for pair production of these negative-$M$ objects, and so some kind of cutoff would be required
for this to be viable\footnote{Invoking a positive-energy criterion asymptotically
would remove these objects from consideration, though it is not clear how to naturally carry this out in the context of the model.}.
We leave these issues for future study.

\subsection{Contribution to the entropy} 

All the above instantons are cosmological solutions. They do not enjoy the presence of an asymptotic region in the Euclidean section or equivalently the presence of a point at infinity. Physically this means these instantons describe closed systems, i.e. systems able to exchange heat and energy. Therefore they necessarily have fixed energy and contribute to the microcanonical ensemble. In such a case it has been shown that the partition function is given by $Z=\Psi^2$, namely as the density of states \cite{Hawking:1995ap,Brown:1992bq,Brown:1994gs}.

As a consequence the contribution to the entropy from these instantons is just
\be
S=\ln Z=-2I_{\mathrm{obj}}.
\ee
For the lukewarm, cold (and Nariai) cases the entropy can be easily obtained from the above formula. As a consistency check for topologies without horizons, the instanton (and consequently the entropy) vanish as expected. We note that horizons contribute to the gravitational entropy only if they contribute to the instanton. Thus  extreme black hole horizons make no contribution to the entropy, even if they have non-zero area. This confirms that the pair creation probability of extreme black holes is lower than that of non-extreme black holes. We can estimate the suppression the pair creation of extreme black holes relative to that of non-extreme black holes as a factor $e^{S_{\mathrm{bh}}}$, where $S_{\mathrm{bh}}$ is the entropy associated with the black hole horizon.

\section{Final remarks \label{concl}}

In this paper we have studied potential quantum instabilities of deSitter spacetime due to nucleation of NCBHs.  We have solved Einstein equations with both nonvanishing cosmological constant and energy momentum tensor.
In order to determine the NCSChwdS solution, the energy momentum tensor was chosen in agreement with smearing prescriptions which led to the NCSchw solution in asymptotically flat space.
We have found an everywhere regular geometry for both positive and negative mass parameter.
For positive masses, the solution admits one, two or three horizons. The latter case corresponds to an inner black hole horizon $r_-$, an outer  black hole horizon $r_+$ and a cosmological horizon $r_c$. In the case of two horizons, $r_-$ and $r_+$ or either $r_+$ and $r_c$ coalesce, corresponding to the case of extremal or Nariai regular black holes. In the case of a single horizon $r_1$ there exists several geometries with positive as well negative mass parameter. These geometries turn out to be topologically equivalent to that with a cosmological horizon, regardless of the value of $r_1$.   On the thermodynamic side, we have computed the temperature of NCSchwdS black holes. Contrary to conventional Schwarzschild black holes, the profile of the temperature admits a maximum value avoiding the divergence in the final stage of the evaporation. Indeed the black hole, after the temperature peak cools down towards a configuration of thermal equilibrium with the deSitter background bath. This final configuration corresponds to the case of a degenerate black hole horizon $r_-=r_+$.  Apart from small deviations due to the deSitter influence, the process resembles what happens in the case of asymptotically flat space NCSchw black holes. On the other hand, the temperature of the cosmological horizon $T_c$ approaches the conventional value $T_{dS}$, even if we always have  $T_c>T_{dS}$ for the thermal contribution of the black hole.  As a second step, we determined the action generating the energy momentum tensor for the NCSchwdS solution. The Euclidean version of this action permits us to analyze the quantum probability of having each of all the aforementioned gravitational objects, namely universes with three, two or one horizon. This probability is compared with that for the deSitter background. By calculating instantons, we shown that the probability of producing gravitational NCBHs or the other gravitational objects is is exponentially small and let us conclude that deSitter space is at the present time quantum mechanically stable in agreement with experience. However we found that the nucleation of NCBHs is relevant for Planckian values of the cosmological constant, i.e. $\Lambda G\sim 1$. Therefore the only time when black hole pair creation was possible in our universe was during the
inflationary era, since during both the radiation and matter dominated eras until the present time, the effective cosmological constant was nearly zero.   In the absence of additional constraints there appears to be an instability toward production of $M<0$ solitons. 

However even for $\Lambda G\sim 1$ the   production of $M>0$ Planck size black holes (or solitons)  seems to be strongly disfavoured, a fact that is against the conventional scenario based on the Schwarzschild-deSitter spacetime \cite{Bousso:1996au}. 
For this reason we believe that the present analysis should be extended to the case of a specific model for inflation, including  recent proposals which exploit, in place of the inflaton field, noncommutative quantum fluctuations to drive the universe expansion  \cite{Rinaldi:2009ba,Garattini:2010dn}. 




\begin{acknowledgements}
This work is supported by the Helmholtz International Center for FAIR within the
framework of the LOEWE program (Landesoffensive zur Entwicklung Wissenschaftlich-\"{O}konomischer
Exzellenz) launched by the State of Hesse and by the Natural Sciences and Engineering Research Council of
Canada. The authors thank the Perimeter Institute
for Theoretical Physics where this work was initiated. The authors thank A. Orlandi for
pointing out a typesetting inaccuracy in the manuscript.

\end{acknowledgements}

\end{document}